\documentclass[twocolumn,secnumarabic,amssymb, nobibnotes, aps, prd, showpacs, showkeys]{revtex4-1}

\usepackage{epsfig}

\begin{document}

\title{Ground state, magnetization process and magnetocaloric effect \\
of the exactly tractable spin-electron tetrahedral chain}%

\author{L. G\'{a}lisov\'{a}$^{1,}$}

\email[e-mail:]{galisova.lucia@gmail.com} 
 
\author{J. Stre\v{c}ka$^2$}%

\affiliation{$^1$Department of Applied Mathematics and Informatics,
             Faculty of Mechanical Engineering, Technical University,
             Letn\'{a} 9, 042 00 Ko\v{s}ice, Slovak Republic
             \\
             $^2$Department of Theoretical Physics and Astrophysics,
             Faculty of Science, P.~J.~\v{S}af\'{a}rik University,
             Park Angelinum 9, 040 01 Ko\v{s}ice, Slovak Republic}%

\begin{abstract}
A hybrid spin-electron system on one-dimensional tetrahedral chain, in which the localized Ising spin regularly alternates with the mobile electron delocalized over three lattice sites, is exactly investigated using the generalized decoration-iteration transformation. The system exhibits either the ferromagnetic or antiferromagnetic ground state depending on whether the ferromagnetic or antiferromagnetic interaction between the Ising spins and mobile electrons is considered. The enhanced magnetocaloric effect during the adiabatic demagnetization suggests a potential use of the investigated system for low-temperature magnetic refrigeration.
\end{abstract}



\date{\today}%
\maketitle

\section{Introduction}
One-dimensional magnetic systems exhibit a variety of interesting phenomena, which are the subject of intensive theoretical and experimental studies nowadays. One of such systems is the coupled tetrahedral chain, which can be observed in the compound Cu$_3$Mo$_2$O$_9$~\cite{1}. The tetrahedral chain structure was first theoretically introduced by Mambrini {\it et al.}~\cite{2} to investigate the residual entropy and spin gap in tetrahedral coupled Heisenberg chain. Since that time, several other models with this lattice geometry have been discussed in the literature, as for instance the Heisenberg and Hubbard models~\cite{3}, the exactly solvable spinless fermion model~\cite{4} and the exactly solvable Ising-Heisenberg model~\cite{5}.

In this work, we will consider another exactly solvable tetrahedral chain, in which the Ising spins localized at nodal lattice sites regularly alternate with triangular plaquettes available for mobile electrons. Assuming the one mobile electron per triangular plaquette, we will study the ground state, the magnetization process and the magnetocaloric effect of this model.

\section{Model and solution}
\label{sec:model}

\begin{figure}[ht]
\begin{center}
\hspace{0.25cm}
\includegraphics[angle = 0, width = 0.8\columnwidth]{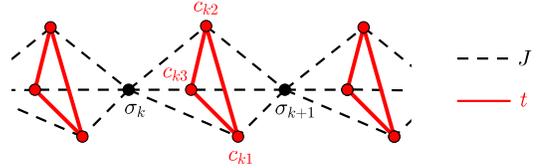}
\vspace{-0.1cm}
\caption{\small A part of the spin-electron system on tetrahedral chain. The black (red) circles indicate the lattice sites available for localized Ising spins (mobile electrons).}
\label{fig1}
\end{center}
\vspace{-0.75cm}
\end{figure}

Let us consider an one-dimensional tetrahedral chain, in which one Ising spin localized at the nodal lattice site regularly alternates with three equivalent lattice sites available for one mobile electron (Fig.~\ref{fig1}). The considered model can be alternatively viewed as the Ising chain, whose bonds are decorated by triangular plaquettes available to mobile electrons. From this perspective, the total Hamiltonian of this model can be written as a sum over cluster Hamiltonians ${\cal H} = \sum_{k =
1}^N{\cal H}_k$, where $N$ is the total number of nodal lattice sites and
the Hamiltonian ${\cal H}_k$  reads
\begin{eqnarray}
{\cal H}_k &=& -t\!\!\sum_{\alpha = \uparrow, \downarrow}\!\!\left(c_{k1,\alpha}^{\dag}c_{k2,\alpha} + c_{k2,\alpha}^{\dag}c_{k3,\alpha} + c_{k3,\alpha}^{\dag}c_{k1,\alpha} + {\rm h.c.}\right) \nonumber\\
&+& \frac{J}{2}(\sigma_{k}^{z} + \sigma_{k+1}^{z})\!\sum_{i = 1}^{3}\!\left(c_{ki,\uparrow}^{\dag}c_{ki,\uparrow} - c_{ki,\downarrow}^{\dag}c_{ki,\downarrow}\right) \nonumber\\
&-& \frac{H_{\rm I}}{2}(\sigma_{k}^{z} + \sigma_{k+1}^{z}) - \frac{H_{\rm e}}{2}\!\sum_{i = 1}^{3}\!\left(c_{ki,\uparrow}^{\dag}c_{ki,\uparrow} - c_{ki,\downarrow}^{\dag}c_{ki,\downarrow}\right)\!.
\label{eq:Hk}
\end{eqnarray}
Here, $c_{ki,\alpha}^{\dag}$ and $c_{ki,\alpha}$ ($\alpha=\uparrow, \downarrow$) represent fermionic creation and annihilation operators, respectively, while $\sigma_{k}^{z}=\pm1/2$ stands for the Ising spins. The parameter $t>0$ takes into account the kinetic energy of a single mobile electron delocalized over a triangular plaquette, the Ising coupling $J$ denotes the exchange interaction between electrons and their nearest Ising neighbours, $H_{\rm I}$ and $H_{\rm e}$ are the Zeeman's terms acting on the localized Ising spins and mobile electrons, respectively. Following the rigorous procedure developed in Ref.~\cite{6} for similar two-dimensional model on doubly decorated planar lattices with one mobile electron delocalized over each couple of decorating sites, one obtains a simple mapping relation between the partition function ${\cal Z}$ of the spin-electron tetrahedral chain and the partition function ${\cal Z}_{\rm IC}$ of the spin-$1/2$ Ising linear chain with the effective nearest-neighbour coupling $J_{e\!f\!f}$ and the effective magnetic field $H_{e\!f\!f}$,
\begin{eqnarray}
{\cal Z}(\beta, J, t, H_{\rm I}, H_{\rm e})= A^N{\cal Z}_{\rm IC}(\beta, J_{e\!f\!f}, H_{e\!f\!f}).
\label{eq:Z}
\end{eqnarray}
In above, $\beta=1/T$ is the inverse temperature (we set $k_{\rm B}=1$) and the unknown parameters $A$,  $J_{e\!f\!f}$ and  $H_{e\!f\!f}$ are given by the expressions
\begin{eqnarray}
A&=& 2\left[2\exp(-\beta t) + \exp(2\beta t)\right]\left(V_-V_+V^2\right)^{1/4}\!,\nonumber\\
J_{e\!f\!f}&=& T\ln\!\left(\frac{V_-V_+}{V^2}\right)\!,\,\,\,\,\, H_{e\!f\!f} = H_{\rm I} + T\ln\!\left(\frac{V_-}{V_+}\right)\!,
\label{eq:ARH}
\end{eqnarray}
where $V_{\mp}=\cosh(\beta J/2 \mp \beta H_{\rm e}/2)$ and $V=\cosh(\beta H_{\rm e}/2)$.
Note that Eq.~(\ref{eq:Z}) allows an exact enumeration of the Helmholtz free energy ${\cal F}=-T\ln{\cal Z}$ of the considered spin-electron tetrahedral chain and, subsequently, other thermodynamic quantities, e.g., the sublattice magnetization $m_{\rm I}=-\frac{1}{N}\!\left(\frac{\partial {\cal F}}{\partial H_{\rm I}}\right)$,  $m_{\rm e}=-\frac{1}{N}\!\left(\frac{\partial {\cal F}}{\partial H_{\rm e}}\right)$ of the localized Ising spins and mobile electrons, respectively, or the entropy per one magnetic particle $s=-\frac{1}{2N}\!\left(\frac{\partial {\cal F}}{\partial T}\right)$.

\section{Results and discussion}
\label{results}
Now, let us proceed to a discussion of the ground state, magnetization process and  magnetocaloric effect of the spin-electron tetrahedral chain with the ferromagnetic ($J<0$) or the antiferromagnetic ($J>0$) Ising interaction. To reduce the number of free interaction parameters, we will assume equal magnetic fields acting on the Ising spins and mobile electrons $H = H_{\rm I}= H_{\rm e}$.

As expected, the investigated tetrahedral chain exhibits the trivial ferromagnetic (FM) ground-state ordering with a full alignment of all nodal Ising spins and mobile electrons into the magnetic field direction, if $J<0$, or it passes from two-fold degenerate antiferromagnetic (AF) ground state to the ferromagnetic (FM) one at the critical field $H_c/|J| =1$, if $J<0$. These two ground states can be characterized through the following eigenvectors and energies per one nodal lattice site:
\begin{eqnarray}
\label{eq:F}
|{\rm FM}\rangle &=& \prod_{k=1}^N
|\!\uparrow\rangle_{\sigma_k}\frac{\displaystyle 1}{\displaystyle \sqrt{3}}\,
(c_{k1,\uparrow}^{\dag}+c_{k2,\uparrow}^{\dag}+c_{k3,\uparrow}^{\dag})|0\rangle_k, \nonumber\\
E_{\rm FM}&=& \frac{J}{2} - 2t-H  ;
\\
\label{eq:AF}
|{\rm AF}\rangle &=& \prod_{k=1}^N\bigg\{
\begin{array}{l}
|\!\uparrow\rangle_{\sigma_k}\frac{\displaystyle 1}{\displaystyle \sqrt{3}}\,
(c_{k1,\downarrow}^{\dag}+c_{k2,\downarrow}^{\dag}+c_{k3,\downarrow}^{\dag})|0\rangle_k
\\[3mm]
|\!\downarrow\rangle_{\sigma_k}\frac{\displaystyle 1}{\displaystyle \sqrt{3}}\,
(c_{k1,\uparrow}^{\dag}+c_{k2,\uparrow}^{\dag}+c_{k3,\uparrow}^{\dag})|0\rangle_k,
\end{array}
\nonumber \\
E_{\rm AF}&=& \frac{J}{2} - 2t.
\end{eqnarray}
In above, the product runs over all primitive unit cells, the ket vector $|\!\uparrow\rangle_{\sigma_k}$ ($|\!\downarrow\rangle_{\sigma_k}$) determines the up (down) state of $k$th localized Ising spin and $|0\rangle_k$ labels the vacuum state for the $k$th triangular plaquette occupied by one mobile electron. It is apparent from Eq.~(\ref{eq:AF}) that the two-fold degenerate AF ground state corresponds to the antiferromagnetic order, in which the mobile electrons are aligned in opposite to their Ising neighbours.

\begin{figure}[hb]
\begin{center}
\vspace{-1.0cm}
\includegraphics[angle = 0, width = 1.0\columnwidth]{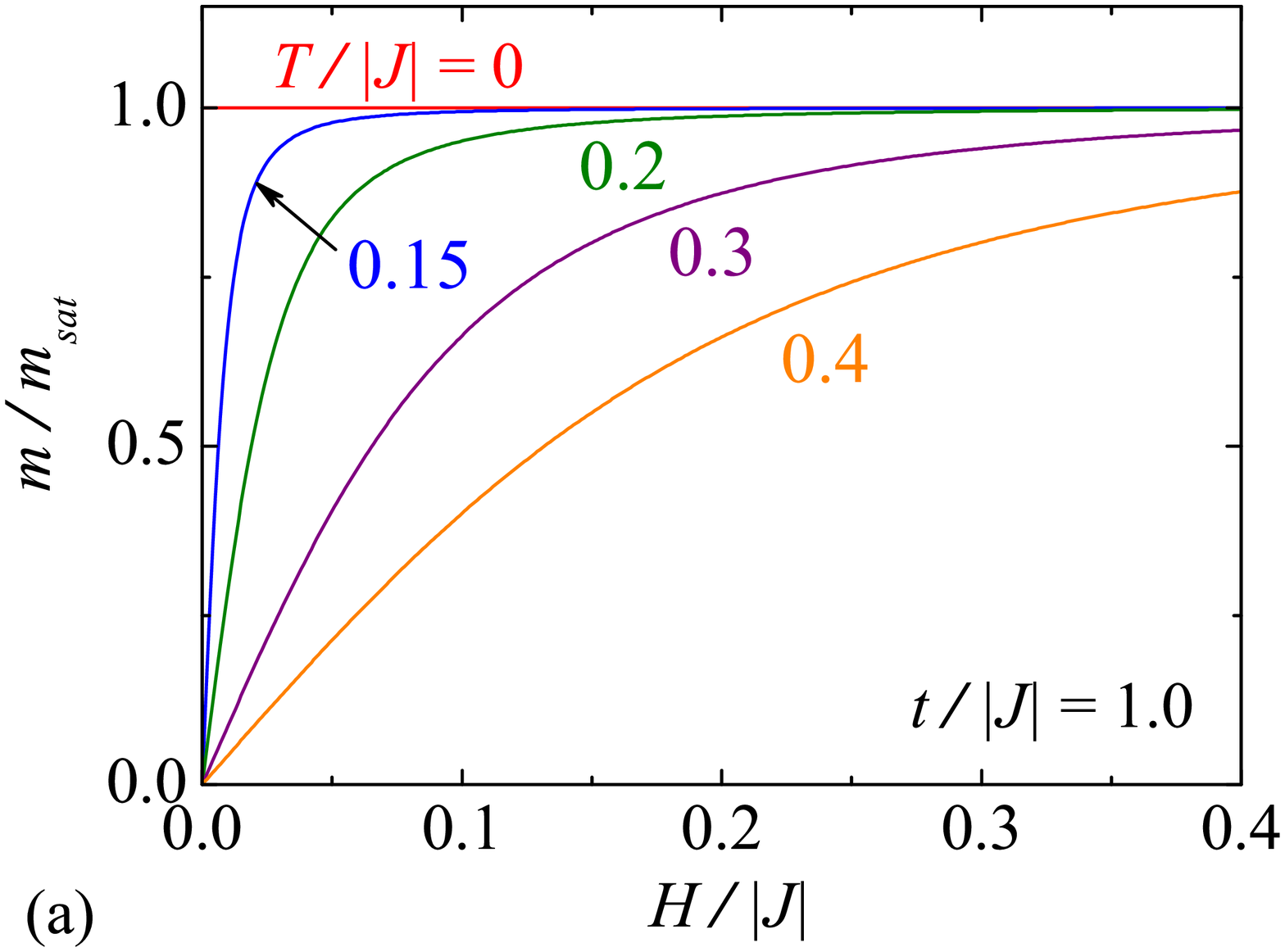}\\[-1.25cm]
\includegraphics[angle = 0, width = 1.0\columnwidth]{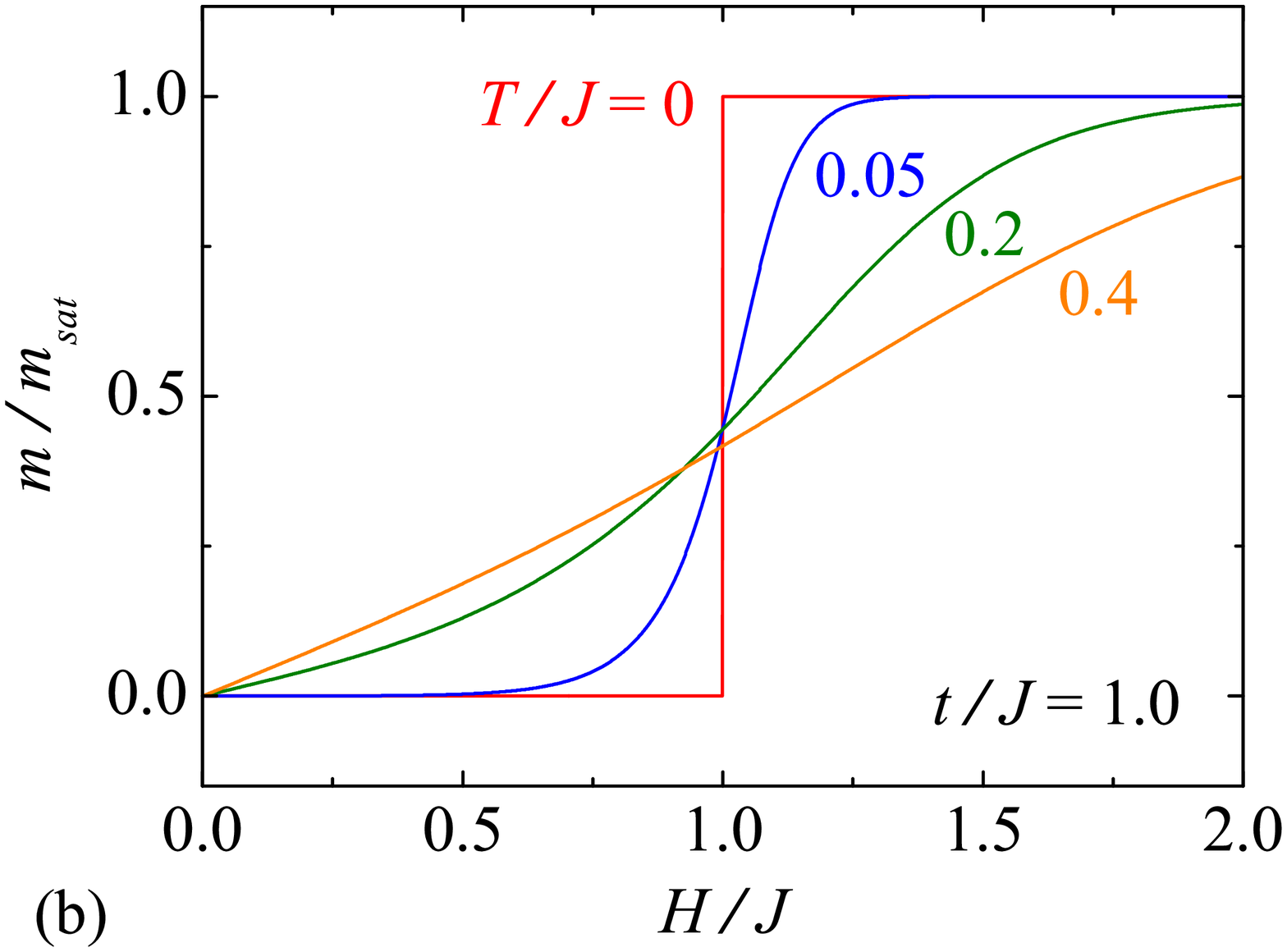}
\vspace{-0.8cm}
\caption{\small Magnetization normalized with respect to its saturation value as a function of the magnetic field at various temperatures for the model with the fixed hopping term $t$ and the Ising interaction: (a)~$J<0$; (b)~$J>0$.}
\label{fig2}
\end{center}
\vspace{-0.25cm}
\end{figure}

Next, let us turn our attention to the magnetic behaviour of the investigated spin-electron model at finite temperatures. In Fig.~\ref{fig2} we depict the magnetization process of the model with the ferromagnetic [Fig.~\ref{fig2}(a)] as well as the antiferromagnetic Ising interaction [Fig.~\ref{fig2}(b)], by assuming the fixed hopping term $t$ and various temperatures. In agreement with the aforedescribed ground-state analysis, the total magnetization [$m=(m_{\rm I}+ m_{\rm e})/2$] of the system with the ferromagnetic Ising interaction $J<0$ takes the saturated value $m_{sat}=1/2$ in the whole field region at $T/J=0$, since all mobile electrons as well as Ising spins are fully polarized into the magnetic field direction. On the other hand, the zero-temperature magnetization curve corresponding to the system with the antiferromagnetic Ising interaction $J>0$ exhibits the abrupt jump from zero to the saturation magnetization at the critical field $H_c/|J| =1$, which coincides with the field-induced phase transition between two possible ground states AF and FM. At any finite temperature, both magnetization curves start from zero in accordance with the one-dimensional character of the investigated spin-electron system. As usual, abrupt changes in low-temperature magnetization curves observable around the fields $H/J =0$ for $J<0$ and $H_c/|J| =1$ for $J>0$ are gradually smeared out upon increasing temperature [see the curves corresponding to $T/J =0.15, 0.2, 0.3, 0.4$ in Fig.~\ref{fig2}(a) and $T/|J| =0.05, 0.2, 0.4$ in Fig.~\ref{fig2}(b)].

\begin{figure}[hb]
\begin{center}
\vspace{-1.0cm}
\includegraphics[angle = 0, width = 1.0\columnwidth]{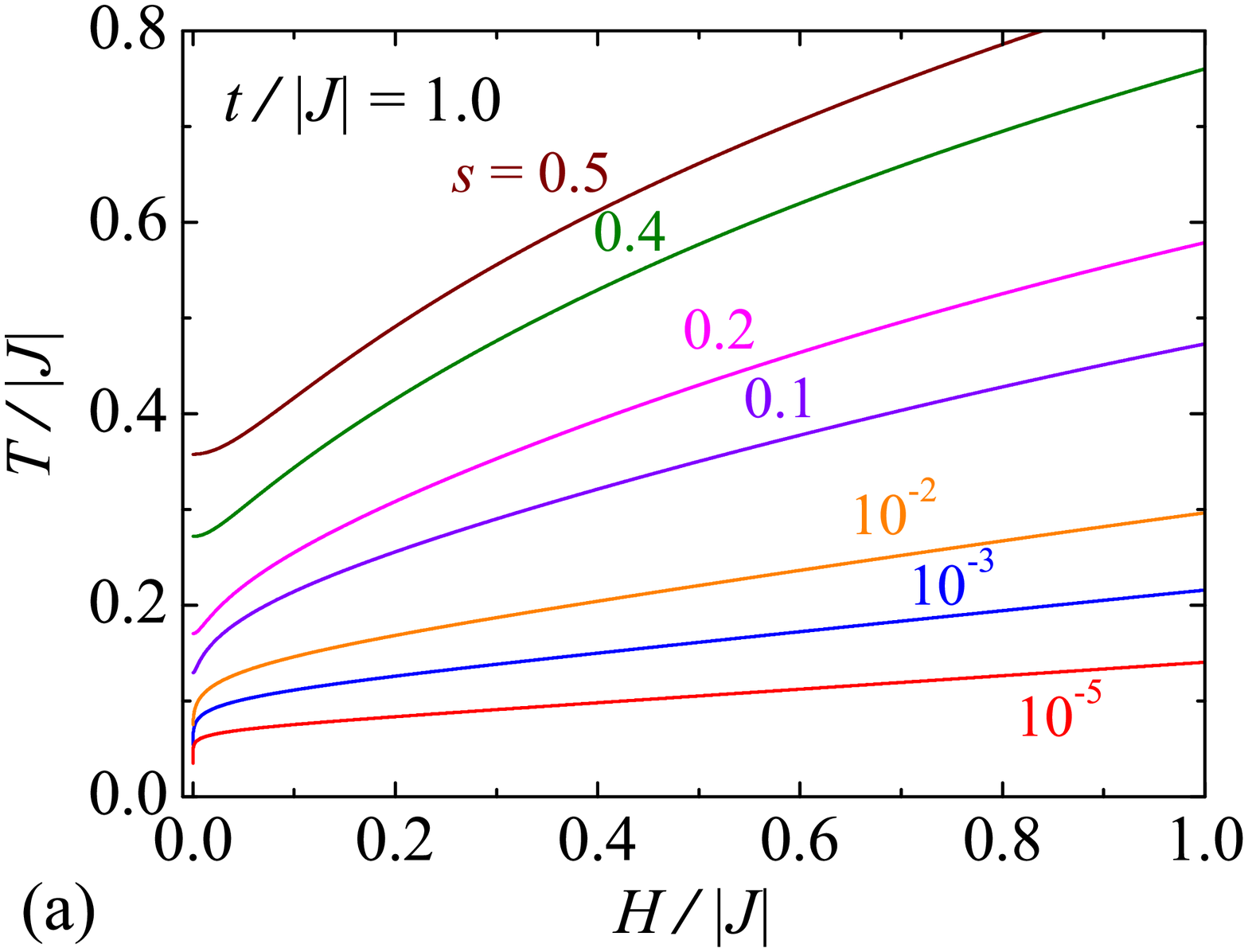}\\[-1.25cm]
\includegraphics[angle = 0, width = 1.0\columnwidth]{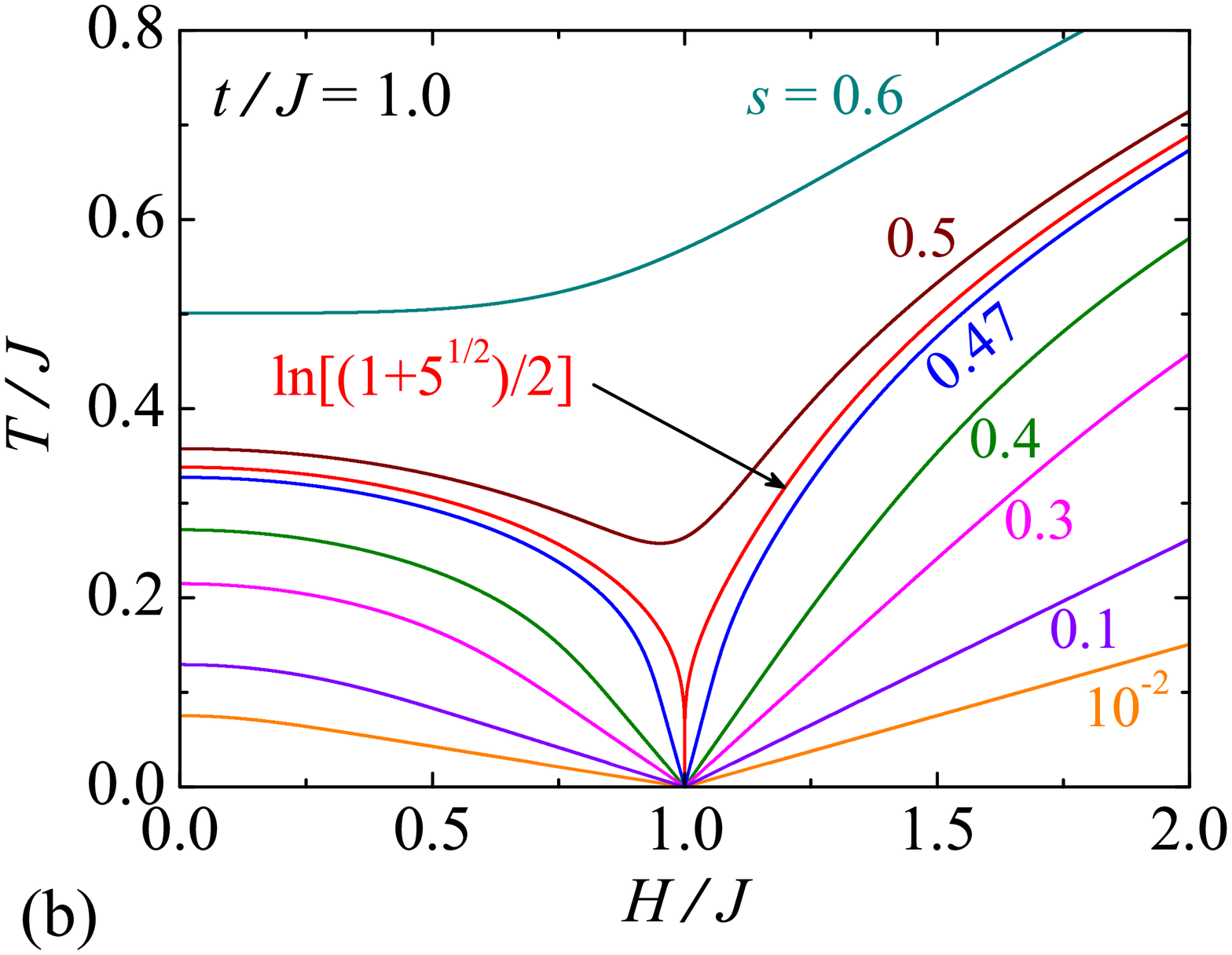}
\vspace{-0.8cm}
\caption{\small Temperature as a function of the magnetic field at various values of the entropy $s$ for the model with the fixed hopping term $t$ and the Ising interaction: (a)~$J<0$; (b)~$J>0$.}
\label{fig3}
\end{center}
\vspace{-0.25cm}
\end{figure}

Finally, let us take a look at the adiabatic demagnetization of the investigated spin-electron model. Fig.~\ref{fig3} shows several typical isentropic changes of the temperature upon varying the external magnetic field for both particular cases with the ferromagnetic and antiferromagnetic exchange coupling $J$. As can be seen from Fig.~\ref{fig3}(a), the model with $J<0$ exhibits an enhanced magnetocaloric effect just around the zero field on assumption that the entropy is very small. By contrast, for other particular case with $J>0$ shown in Fig.~\ref{fig3}(b), the most obvious drop/grow of the temperature can be found in the vicinity of the critical field $H_c/J=1$, where the system undergoes a zero-temperature phase transition between the AF and FM ground states. Obviously, an enhanced magnetocaloric effect can be detected in this particular case just if the entropy is sufficiently close to the value $s=\ln\left(\frac{1+\sqrt{5}}{2}\right)$.

\begin{figure}[h]
\begin{center}
\vspace{-0.5cm}
\includegraphics[angle = 0, width = 1.05\columnwidth]{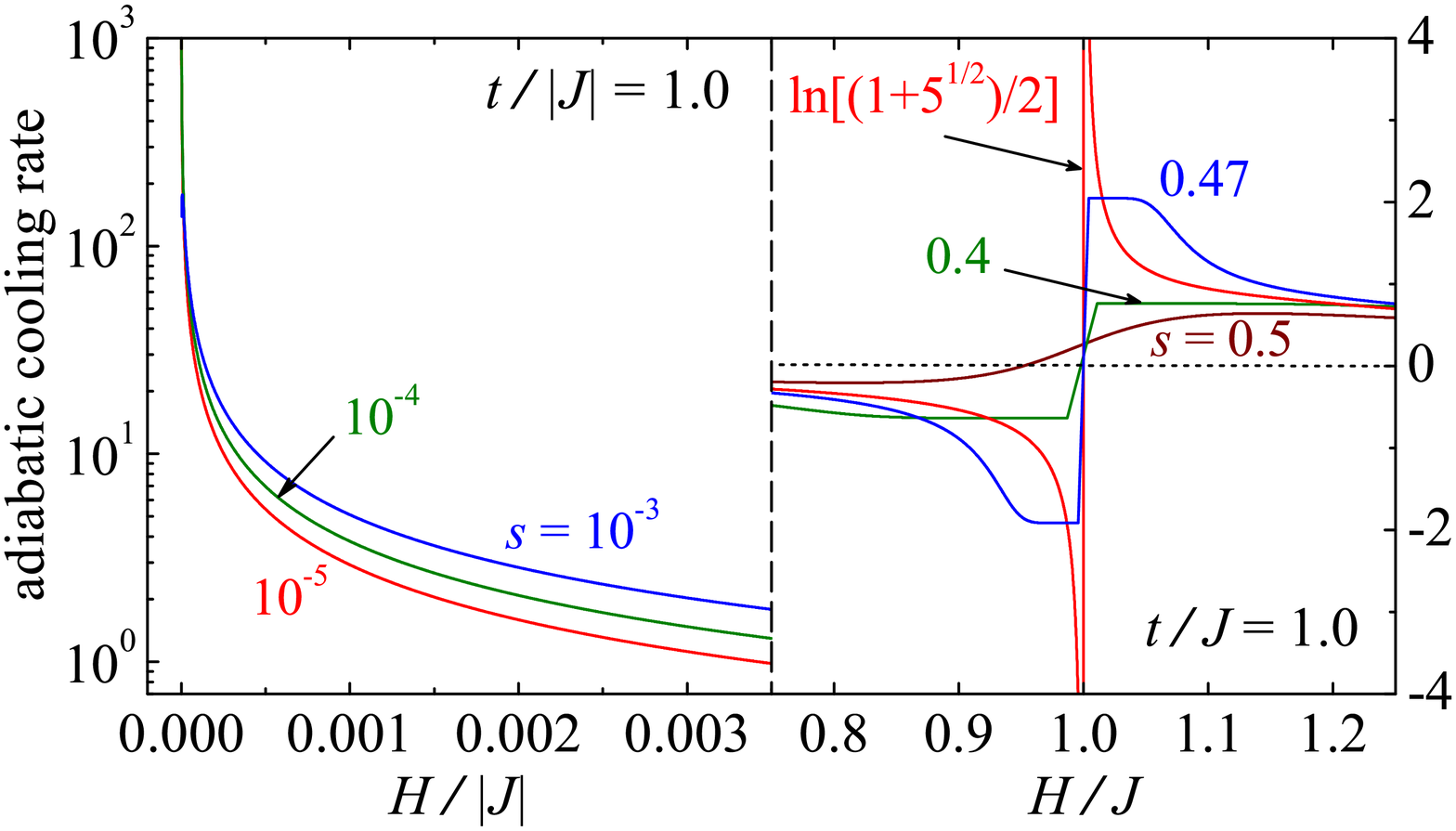}
\vspace{-1.0cm}
\caption{\small Adiabatic cooling rate as a function of the magnetic field for the model with the fixed hopping term $t$ and the Ising interaction $J<0$ (left panel), $J>0$ (right panel) at various values of the entropy.}
\label{fig4}
\end{center}
\vspace{-0.5cm}
\end{figure}

To describe an efficiency of the cooling process in the studied model,
the adiabatic cooling rate $\left(\frac{\partial T}{\partial H}\right)_s$
as a function of the external magnetic field is illustrated in Fig.~\ref{fig4} for both particular cases $J>0$ and $J<0$ by keeping the entropy constant. In agreement with the previous analysis, the model with $J<0$ shows the asymptotically fast cooling during the adiabatic demagnetization around $H/|J|=0$ when the entropy is small enough, whereas the model with $J>0$ exhibits the most rapid cooling/heating during the adiabatic demagnetization just around the transition field $H_c/J=1$ when the entropy is sufficiently close to the value $s=\ln\left(\frac{1+\sqrt{5}}{2}\right)$.

\section{Summary}
\label{summary}
In this paper, we have studied the ground state, the magnetization process and the enhanced magnetocaloric effect of the exactly solvable spin-electron tetrahedral chain by assuming one Ising spin localized at each nodal lattice site and one mobile electron delocalized over each triangular plaquette. It has been demonstrated that the system with the ferromagnetic  interaction $J<0$ exhibits an enhanced magnetocaloric effect when the entropy is kept very small, while the system with the antiferromagnetic interaction $J>0$ exhibits an enhanced magnetocaloric effect when the entropy is set sufficiently close to its optimal value $s=\ln\left(\frac{1+\sqrt{5}}{2}\right)$. This suggests a potential use of the investigated system for low-temperature magnetic refrigeration.
\\[4mm]
{\bf Acknowledgments}:
This work was financially supported by the grant of the Slovak Research and Development Agency under the contract No. APVV-0097-12.


\begin{thebibliography}{6}
\bibitem{1} M. Hase, H. Kitazawa, K. Ozawa, T. Hamasaki, H.
Kuroe, T. Sekine, J. Phys. Soc. Jpn. {\bf 77}, 034706 (2008).
\bibitem{2} M. Mambrini, J. Tr\'{e}bosc, and F. Mila, Phys. Rev. B {\bf 59}, 13806 (1999).
\bibitem{3} M. Maksymenko, O. Derzhko, and J. Richter, Eur. Phys. J. B {\bf 84}, 397 (2011).
\bibitem{4} M. Rojas, S. M. de Souza, O. Rojas, arXiv:1212.5552.
\bibitem{5} D. Antonosyan, S. Bellucci and V. Ohanyan, Phys. Rev. B {\bf 79}, 014432 (2009).
\bibitem{6} J. Stre\v{c}ka, A. Tanaka, M. Ja\v{s}\v{c}ur,  J. Phys.: Conf. Ser. {\bf 200} 022059 (2010).
\end{thebibliography}
\end{document}